\documentclass[aps,twocolumn,prl,showpacs,preprintnumbers,superscriptaddress,amsmath,amssymb,floatfix]{revtex4}
\usepackage{graphicx}
\usepackage{dcolumn}
\usepackage{bm}
\usepackage[english]{babel}

\begin{document}
\preprint{APS/123-QED}

\author{R. K\"uchler}
\author{P. Gegenwart}
\author{J. Custers}\thanks{present address:
 Institute of Solid State Physics, Vienna University of Technology,
  1040  Vienna, Austria}
\author{O. Stockert}
\author{N. Caroca-Canales}
\author{C. Geibel}
\affiliation{Max-Planck-Institute for Chemical Physics of Solids,
D-01187 Dresden, Germany}
\author{J.G. Sereni}
\affiliation{Lab. Bajas Temperaturas, Centro Atomico Bariloche, 8400
S.C. Bariloche, Argentina}
\author{F. Steglich}
\affiliation{Max-Planck-Institute for Chemical Physics of Solids,
D-01187 Dresden, Germany}

\title{Quantum criticality in the cubic heavy-fermion system
CeIn$_{3-x}$Sn$_x$}

\date{
\today%
}

\begin{abstract}
We report a comprehensive study of CeIn$_{3-x}$Sn$_x$ $(0.55 \leq
x \leq 0.8)$ single crystals close to the antiferromagnetic (AF)
quantum critical point (QCP) at $x_c\approx 0.67$ by means of the
low-temperature thermal expansion and Gr\"uneisen parameter. This
system represents the first example for a {\it cubic} heavy
fermion (HF) in which $T_{\rm N}$ can be suppressed {\it
continuously} down to $T=0$. A characteristic sign change of the
Gr\"uneisen parameter between the AF and paramagnetic state
indicates the accumulation of entropy close to the QCP. The
observed quantum critical behavior is compatible with the
predictions of the itinerant theory for three-dimensional critical
spinfluctuations. This has important implications for the role of
the dimensionality in HF QCPs.

\end{abstract}

\pacs{71.10.HF,71.27.+a} \maketitle

Non-Fermi-liquid (NFL) properties are observed in many
heavy-fermion (HF) systems and frequently attributed to a nearby
quantum critical point (QCP) \cite{Sachdev-book}. A QCP can arise
by continuously suppressing the transition temperature $T_{\rm
N}$ of an antiferromagnetic (AF) phase to zero, e.g. by chemical
or applied pressure or an external magnetic field. QCPs are of
great current interest due to their singular ability to influence
the finite temperature properties of materials. Heavy-fermion
metals have played the key role in the study of AF QCPs. The
essential question is how the heavy quasiparticles evolve if these
materials are tuned from the paramagnetic into the AF ordered
state. The traditional picture describes a spin-density-wave (SDW)
transition and related, a mean-field type of quantum critical
behavior. Here, the quasiparticles retain their itinerant
character \cite{Millis,Moriya}. Unconventional quantum
criticality which qualitatively differs from the standard theory
of the $T=0$ SDW transition, may arise due to a destruction of
Kondo screening. Here, the quasiparticles break up into their
components: conduction electrons and local 4$f$ moments forming
magnetic order \cite{Si,Coleman}. This locally-critical picture
leads to a number of distinct properties, including stronger than
logarithmic mass divergence,
$\omega/T$ scaling in the dynamical susceptibility and a large
reconstruction of the Fermi surface. Such behavior has been found
at least in some HF systems \cite{Schroeder,Custers,Paschen}. The
central question is to identify the crucial parameter leading to
the different types of QCPs. Of particular importance should be
the dimensionality of the magnetic fluctuations, which could be
reduced by the presence of frustration.
It is proposed in \cite{Si} that for magnetically
three-dimensional (3D) systems without frustration the itinerant
SDW picture should apply. On the other hand, 2D magnetic systems
should be described by a locally quantum critical picture
\cite{Si}. However, systems currently under investigation, are
either tetragonal, e.g. CeNi$_2$Ge$_2$,
YbRh$_2$(Si$_{1-x}$Ge$_x$)$_2$ \cite{Kuechler} and CeCu$_2$Si$_2$
\cite{Gegenwart}, hexagonal, e.g. YbAgGe \cite{Budko} or
monoclinic, e.g. CeCu$_{6-x}$Au$_{x}$ \cite{Schroeder} and the
lower crystallographic symmetry could result in fluctuations with
reduced dimensionality. Therefore the dimensionality of the
critical spin fluctuations clearly needs to be substantiated by
inelastic neutron scattering experiments. In order to avoid this
constraint, experiments on cubic systems close to QCPs are
particular interesting. CeIn$_{3-x}$Sn$_x$, with a cubic point
symmetry of Ce-atoms in the Cu$_3$Au structure (compare the inset
of Figure 1), is thus an excellent candidate for such a study, as
here low-dimensional spin fluctuations can be ruled out. Thus,
the interesting question arises, whether the mechanism of
NFL-behavior in this system can be described by an itinerant 3D
SDW theory.

In this Letter, we present thermal expansion measurements and a
Gr\"uneisen ratio analysis performed on single crystalline samples
of the cubic system CeIn$_{3-x}$Sn$_x$ close to the critical
concentration, $x_c = 0.67$, where $T_{\rm N}$ is suppressed to
zero by doping. Recently, it has been shown that the thermal
volume expansion $\beta=V^{-1}(dV/dT)$ ($V$: sample volume) is
particular suited to probe quantum-critical behavior, since,
compared to the specific heat, it is much more singular in the
approach to the QCP \cite{Zhu}. As a consequence, the Gr\"uneisen
ratio $\Gamma \sim \beta/C$ of thermal expansion, $\beta(T)$, to
specific heat, $C(T)$, is divergent as $T$ goes to zero at any
pressure-sensitive QCP and the associated critical exponent can
be used to distinguish between the different types of QCPs. In
the itinerant scenario the divergence $\Gamma\propto
1/T^\epsilon$ is given by $\epsilon=1/\nu z$ \cite{Zhu} with
$\nu$, the critical exponent for the correlation length,
$\xi\propto |r|^\nu$ ($r$: distance from the QCP) and $z$, the
dynamical critical exponent in the divergence of the correlation
time, $\tau_c\propto\xi^z$. For a 3D AF QCP $\nu=1/2$ and $z=2$
yielding $\epsilon=1$. Thus, a study of the Gr\"uneisen ratio can
prove the validity of the 3D SDW picture in the title system
provided that $\epsilon=1$.

\begin{figure}
\centerline{\includegraphics[width=\linewidth,keepaspectratio]{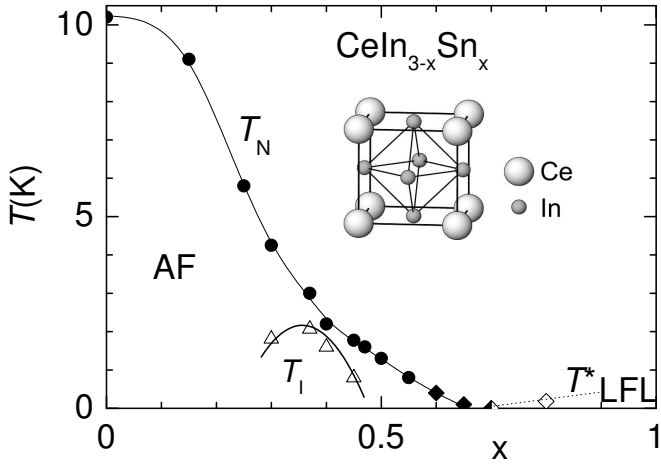}}
\caption{Magnetic phase diagram for cubic CeIn$_{3-x}$Sn$_x$ ($x
\leq 1$). Closed circles and diamonds indicate $T_{\rm N}$,
determined from specific heat \cite{PedrazziniB} and electrical
resisitivity \cite{CustersB} measurements, respectively. Open
diamonds mark $T^{\star}$, the upper limit of Landau Fermi-liquid
behavior, e.g. $\bigtriangleup \rho(T) \propto T^{2}$
\cite{CustersC}. Open triangles indicate first-order transition
$T_I$ \cite{PedrazziniB}.} \label{fig1}
\end{figure}

The magnetic $(x,T)$ phase diagram of polycrystalline
CeIn$_{3-x}$Sn$_x$ has been widely studied for $0\leq x\leq 1$ by
susceptibility \cite{Lawrence}, specific-heat \cite{PedrazziniB} and
resistivity measurements \cite{CustersB,CustersC} (see Figure 1).
Whereas $T_N$ for undoped CeIn$_3$ vanishes discontinuously below
3~K under hydrostatic pressure \cite{Mathur}, it can be traced down
to 0.1~K for CeIn$_{3-x}$Sn$_x$ and an additional first-order phase
transition $T_I$ has been found for $0.25 < x <0.5$ inside the AF
state \cite{PedrazziniB}. These differences are related to the
change of the electronic structure induced by Sn doping. Beyond a
possible tetracritical point at $x\approx 0.4$ \cite{PedrazziniB},
an almost linear dependence of $T_N(x)$ is observed. This is in
contrast to $T_N\propto (x_c-x)^{2/3}$ predicted by the 3D-SDW
theory \cite{Millis}.
Thus, the origin of the NFL behavior in this system remains an open
question, and further thermodynamic studies are needed to shed light
on the nature of the QCP.

The CeIn$_{3-x}$Sn$_x$ single crystals investigated here (0.55
$\leq x \leq 0.80$) were grown by a Bridgman-type technique. Large
single crystals with a mass of 15~g were produced, analyzed by
X-ray powder diffraction and found to be of single phase with the
proper cubic structure. Within the $\pm 2\%$ accuracy of the
X-ray diffraction, no impurity phases were resolvable. Thin bars
with a length 2 $ \leq l \leq 6$ mm, suitable for the
dilatometric investigations, were cut out.
The thermal expansion has been measured in a dilution refrigerator
using an ultrahigh resolution capacitive dilatometer with a
maximum sensitivity corresponding to $\Delta l/l = 10^{-11}$.

\begin{figure}
\centerline{\includegraphics[width=\linewidth,keepaspectratio]{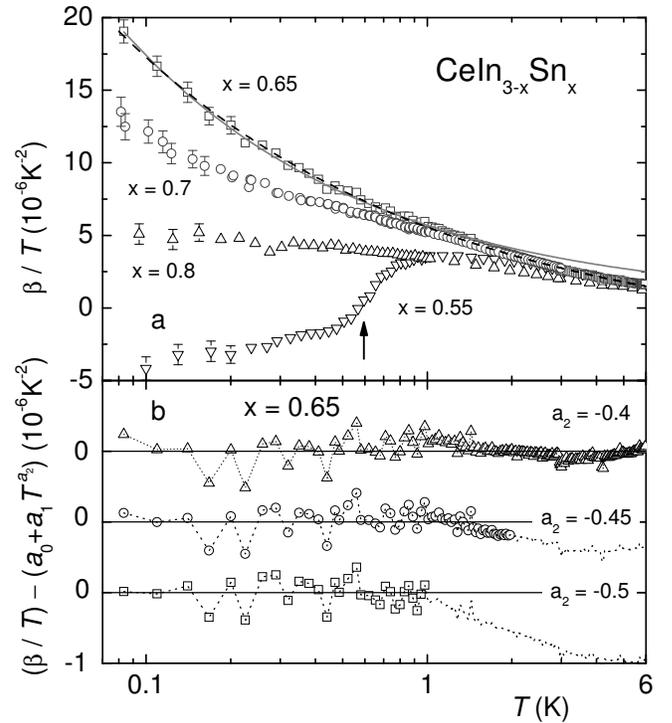}}
\caption{Volume thermal expansion coefficient $\beta$ of
CeIn$_{3-x}$Sn$_x$ single crystals as $\beta/T$ vs $\log T$ (a).
Gray solid and black dashed lines indicate $T^{-0.5}$ and $T^{-0.4}$
dependencies, respectively. Arrow indicates AF phase transition.
(b): Deviation of $\beta/T$ data for $x=0.65$ sample from best
power-law fits for $T\leq 1$~K (squares), $T\leq 2$~K (circles), and
$T\leq 6$~K (triangles), respectively, as
$(\beta/T)-(a_0+a_1T^{a_2})$ vs $\log T$. For clarity, the three
data sets have been shifted by different amounts vertically.}
\label{fig2}
\end{figure}

Fig. 2a shows the volume thermal expansion $\beta$ of
single-crystalline CeIn$_{3-x}$Sn$_x$ with $x=0.55, 0.65, 0.7$ and
0.8 plotted as $\beta(T)/T$ vs $\log T$. The volume-expansion
coefficient $\beta$ is given by $\beta=3\times\alpha$, with $\alpha$
being the linear thermal-expansion coefficient.
For $x=0.55$ the broadened step-like decrease in $\beta /T$ at
$T_{\rm N}\approx 0.6$~K marks the AF phase transition, in
perfect agreement with specific-heat measurements on the same
single crystal \cite{Rus}. Upon increasing the concentration we
find for $x=0.65$ and $0.7$ diverging behavior over nearly two
decades in $T$ down to 80~mK. These data suggest that $T_{\rm N}$
is suppressed at a critical concentration $x_c\approx 0.67\pm
0.03$, also consistent with specific-heat measurements performed
on the same samples \cite{Rus}. Finally, for $x=0.8$ we recover
Fermi-liquid behavior, $\beta(T)/T \approx const.$ for
$T\rightarrow 0$.

In the following, we will analyze the observed NFL behavior and
make comparison with the predictions of the itinerant SDW scenario
\cite{Zhu}. 
A best-fit description of the $x=0.65$ data in the entire
temperature range $0.08$~K$\leq T\leq~6$~K according to
$\beta/T=a_0+a_1T^{a_2}$ reveals $a_2=-0.4\pm 0.01$ (see dashed line
in Fig.~2a). However, as shown in the upper part of Fig.~2b, the
deviation between the data and this fit shows several broad bumps
indicating that the fit does not properly describes the data. We
therefore tried best power-law fits for $0.08$~K$\leq T\leq T_{max}$
with varying $T_{max}$. For $T_{max}=1$~K, the fit is of excellent
quality (cf. solid line in Fig.~2a and lower part of Fig.~2b) and
the resulting exponent equals $-0.5$, i.e. the value predicted by
the 3D SDW scenario \cite{Zhu}.

\begin{figure}
\centerline{\includegraphics[width=0.95\linewidth,keepaspectratio]{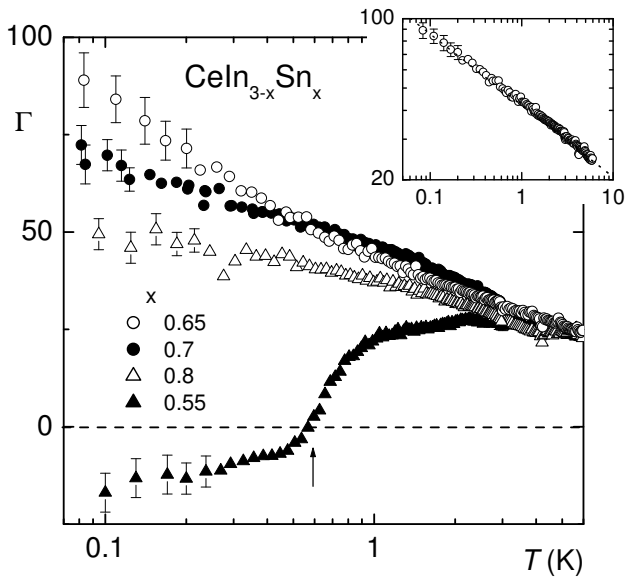}}
\caption{Temperature dependence of the Gr\"uneisen parameter
$\Gamma=V_m/\kappa_T\cdot\beta/C$ of several CeIn$_{3-x}$Sn$_x$
single crystals as $\Gamma(T)$ vs $\log T$. $V_m=6.25\cdot 10^{-5}$
m$^3$mol$^{-1}$ and $\kappa_T=1.49\cdot 10^{-11}$ Pa$^{-1}$
\cite{Vedel} are the molar volume and isothermal compressibility,
respectively. Arrow indicates AF phase transition. The inset
displays data for $x=0.65$ in a double-logarithmic plot. The dotted
line indicates the power-law dependence $\Gamma\propto T^{-0.31}$.}
\label{fig3}
\end{figure}

We now turn to the Gr\"uneisen parameter defined as
$\Gamma=V_m/\kappa_T\cdot\beta/C$ where the constants $V_m$ and
$\kappa_T$ denote the molar volume and isothermal compressibility,
respectively. The specific heat has been studied in the temperature
range 40~mK~$\leq T\leq$~4~K on the same CeIn$_{3-x}$Sn$_x$ samples
used for thermal expansion \cite{Rus,Radu}. Here, the nuclear
quadrupole contribution of indium which becomes important below
150~mK has been subtracted. Fig.~3 shows a comparison of $\Gamma(T)$
for all samples studied in thermal expansion. Since the temperature
dependence of specific heat is much weaker compared to that of
thermal expansion, its influence to the Gr\"uneisen parameter is
rather small. Therefore the variation of $\Gamma(T)$ for the
different CeIn$_{3-x}$Sn$_x$ samples is very similar to that found
in $\beta/T$ (compare Fig.~2). Both single crystals closest to
$x_c$, $x=0.65$ and $x=0.7$ show a divergent behavior down to the
lowest accessible temperature with very large $\Gamma$ values at
0.1~K which are of similar size as found for other quantum critical
HF systems \cite{Kuechler,Kuechler04}. On the other hand, saturation
is observed for $x=0.55$ and $x=0.8$ being located in the AF ordered
and Fermi liquid regime, respectively. The fact that the divergence
of $\Gamma(T)$ in the quantum critical regime is stronger than
logarithmic (compare the double logarithmic representation of the
$x=0.65$ data presented in the inset of Fig.~3) provides clear
evidence for a well defined (pressure-sensitive) QCP in the system.
If the disorder present in the system would lead to a "smeared"
quantum critical regime, $\Gamma(T)$ could diverge at most
logarithmically \cite{Zhu}.

Another indication for a QCP is the sign change of the Gr\"uneisen
parameter between the ordered and disordered regime. As discussed in
\cite{Garst},
it is directly related to the entropy accumulation near the QCP.
The different signs of $\Gamma$ in the AF and paramagnetic region
reflect the opposite pressure dependencies of the respective
characteristic energy scales. Below $T_{\rm N}$, the effective AF
intersite interaction dominates, whose negative pressure dependence
gives rise to a negative Gr\"uneisen parameter, $\Gamma < 0$. On the
other hand, the positive Gr\"uneisen ratio in the paramagnetic state
is compatible with the positive pressure dependence of the Kondo
temperature in Ce-based HF Systems.

\begin{figure}
\centerline{\includegraphics[width=\linewidth,keepaspectratio]{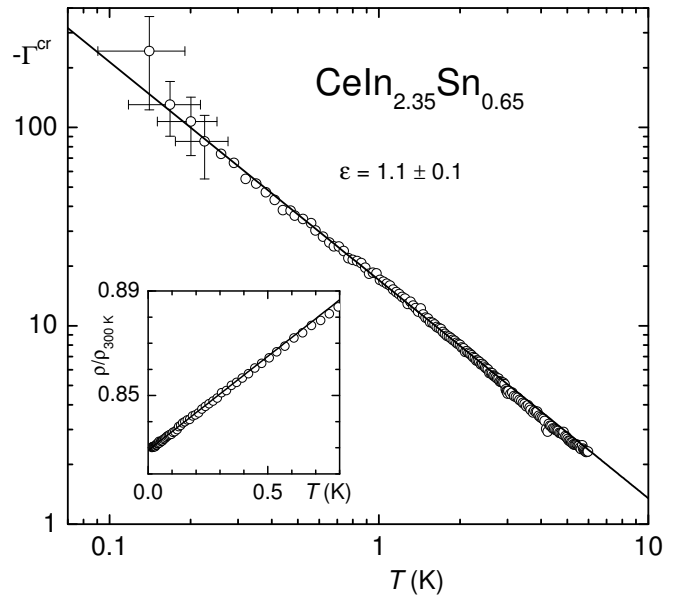}}
\caption{Critical Gr\"uneisen ratio
$\Gamma^{cr}=V_m/\kappa_T\cdot\beta^{cr}/C^{cr}$ for
CeIn$_{2.35}$Sn$_{0.65}$ as $\log\Gamma^{cr}$ {\it vs} $\log T$ with
critical components $\beta^{cr}=\beta(T)-a_0T$ and
$C^{cr}=C(T)-\gamma T$ derived after subtraction of background
contributions (see text). Solid line represents $\Gamma^{cr}\propto
1/T^{\epsilon}$ with $\epsilon=1.1 \pm 0.1$. The inset shows the
low-temperature electrical resistivity of a single crystal of
similar composition.} \label{fig4}
\end{figure}

In order to compare our results for the $x = 0.65$ sample which is
located closest to the QCP with the theoretical predictions for an
itinerant AF QCP \cite{Zhu}, we need to calculate the {\it critical}
Gr\"uneisen ratio $\Gamma^{cr}(T)\propto \beta^{cr}/C^{cr}$ of
critical contributions to thermal expansion and specific heat. For
thermal expansion, $\beta^{cr}(T)=\beta(T)-a_0T$ with $a_0=0.3\times
10^{-6}$~K$^{-2}$ as determined from the best fit up to 1~K, see
above. Within the itinerant theory for the 3D AF case, the critical
contribution to specific heat is {\it sub-leading} \cite{Zhu}:
$C^{cr}(T)=C(T)-\gamma_0T$, with $C^{cr}<0$, $C^{cr}/T\rightarrow 0$
for $T\rightarrow 0$ and $\gamma_0=C/T \mid _{T=0}$. For $\gamma_0$
we use the value 0.851~Jmol$^{-1}$K$^{-2}$ obtained in \cite{Rus}
from fitting the low-temperature electronic specific heat in a {\it
restricted} temperature range 0.3~K~$\leq T\leq$~1.4~K according to
$C/T=\gamma_0(1-a'\sqrt{T})$. Figure 3 displays a log-log plot of
$\Gamma^{cr}(T)$ versus temperature. We find $\Gamma^{cr}\propto
T^{-\epsilon}$ with an exponent $\epsilon=1.1\pm 0.1$ which is very
close to 1, predicted by the itinerant theory. Note, that this
exponent is rather insensitive of $\gamma_0$ subtracted from the
specific heat data: using $\gamma_0=0.9$~Jmol$^{-1}$K$^{-2}$ and
0.95~Jmol$^{-1}$K$^{-2}$ results in $\epsilon=1.07$ and $1.02$,
respectively. Interestingly, the exponent for the critical
Gr\"uneisen ratio, which theoretically equals the dimension of the
most relevant operator that is coupled to pressure \cite{Zhu}, holds
over a much larger temperature range than the respective 3D-SDW
dependencies in specific heat \cite{Rus} and thermal expansion (cf.
Fig. 2). A similar observation has also been made for CeNi$_2$Ge$_2$
\cite{Kuechler}.

For those two systems for which an unconventional QCP has been
proposed, YbRh$_2$Si$_2$ and CeCu$_{6-x}$M$_x$ (M=Au, Ag),
distinctly different temperature dependences have been observed:
$\Gamma^{cr}\propto T^{-0.7}$ in the former \cite{Kuechler} and
$\Gamma^{cr}\propto \log T$ \cite{Kuechler04} in the latter case.
It is proposed in \cite{Si}, that for magnetically 3D systems
without frustration the SDW picture should apply. This is
consistent with our Gr\"uneisen ratio analysis.

For the 3D AF case, the itinerant theory predicts an {\it
asymptotic} $T^{3/2}$ dependence for the temperature dependent part
to the electrical resistivity \cite{Millis,Moriya}. As discussed in
\cite{Rosch}, the interplay between strongly anisotropic scattering
due to the critical spinfluctuations and isotropic impurity
scattering can lead at {\it elevated temperature} to temperature
exponents of the resistivity between 1 and 1.5, depending on the
amount of disorder. Systematic $\rho(T)$ studies down to mK
temperatures on polycrystalline CeIn$_{3-x}$Sn$_x$ revealed an
almost linear temperature dependence in the quantum critical regime
\cite{CustersB}. Similar behavior is observed for single crystalline
CeIn$_{2.35}$Sn$_{0.65}$ as well, see the inset of Figure 4.
However, due to the high Sn-doping needed to tune the system towards
the QCP, the resistivity ratio $\rho_{300 K}/\rho_0$ is of the order
of 1 and the temperature variation amounts to a few percent of
$\rho_0$ only, making the comparison with theoretical predictions
very difficult. This indicates that transport experiments alone are
not sufficient to characterize quantum criticality in disordered
systems. Possibly, also the slope of $T_N(x)$ differs from the
3D-SDW prediction because disorder is not constant but increases
with increasing $x$.
However, the algebraic divergence of $\Gamma(T)$ for $T\rightarrow
0$ at $x\approx x_c$ proves a pressure-sensitive QCP in the system
and excludes disorder-driven scenarios for the observed NFL behavior
\cite{Zhu}.


In conclusion, our study on CeIn$_{3-x}$Sn$_x$ single crystals by
means of the low-temperature thermal expansion and Gr\"uneisen
parameter has proven the applicability of the itinerant theory for
3D critical spinfluctuations in this cubic system. Since strong
contradictions to this theory have been found in systems like
CeCu$_{5.9}$Au$_{0.1}$ \cite{Schroeder} or
YbRh$_2$(Si$_{1-x}$Ge$_x$)$_2$ \cite{Custers} with lower
crystallographic symmetry and, at least in case of the former
system, strongly anisotropic quantum critical fluctuations, the
parameter {\it dimensionality} obviously plays an important role
for the nature of HF QCPs. We tentatively classify the different
HF systems studied by Gr\"uneisen analysis at their respective
QCPs as follows: (i) CeIn$_{3-x}$Sn$_x$ and CeNi$_2$Ge$_2$
\cite{Kuechler} for which latter system neutron scattering
measurements revealed 3D low-energy magnetic fluctuations
\cite{Kadowaki}, show thermodynamic behavior compatible with the
3D itinerant theory, whereas for (ii)
YbRh$_2$(Si$_{1-x}$Ge$_x$)$_2$ \cite{Kuechler} and
CeCu$_{5.8}$Ag$_{0.2}$ \cite{Kuechler04} strong contradictions to
this model (for both 2D and 3D critical spinfluctuations) are
observed. Neutron scattering has proven 2D quantum critical
fluctuations in CeCu$_{5.9}$Au$_{0.1}$ \cite{Stockert} while for
YbRh$_2$Si$_2$ a complicated behavior with competing AF and
ferromagnetic quantum critical fluctuations has been observed 
\cite{Ishida}. 
The comparison with our results on CeIn$_{3-x}$Sn$_x$ suggests that
a destruction of Kondo screening causing unconventional quantum
criticality is prevented in magnetically 3D systems.

We thank T. Radu for providing low-temperature specific heat data.
Work supported in part by the F. Antorchas - DAAD cooperation
program (Pr. Nr. 1428/7).

\end{document}